# Digital twin brain: a bridge between biological intelligence and artificial intelligence

Hui Xiong[1], Congying Chu[1], Lingzhong Fan[12], Ming Song[1], Jiaqi Zhang[12], Yawei Ma[12], Ruonan Zheng[3], Junyang Zhang[3], Zhengyi Yang[1], Tianzi Jiang[123*]

1 Brainnetome Center, Institute of Automation, Chinese Academy of Sciences, 100190, Beijing, China
2 School of Artificial Intelligence, University of Chinese Academy of Sciences, 100049, Beijing, China
3 Research Center for Augmented Intelligence, Zhejiang Lab, 311100, Hangzhou, China
* Corresponding author: jiangtz@nlpr.ia.ac.cn

**Abstract**
In recent years, advances in neuroscience and artificial intelligence have paved the way for unprecedented opportunities for understanding the complexity of the brain and its emulation by computational systems. Cutting-edge advancements in neuroscience research have revealed the intricate relationship between brain structure and function, while the success of artificial neural networks highlights the importance of network architecture. Now is the time to bring them together to better unravel how intelligence emerges from the brain's multiscale repositories. In this review, we propose the Digital Twin Brain (DTB) as a transformative platform that bridges the gap between biological and artificial intelligence. It consists of three core elements: the brain structure that is fundamental to the twinning process, bottom-layer models to generate brain functions, and its wide spectrum of applications. Crucially, brain atlases provide a vital constraint, preserving the brain's network organization within the DTB. Furthermore, we highlight open questions that invite joint efforts from interdisciplinary fields and emphasize the far-reaching implications of the DTB. The DTB can offer unprecedented insights into the emergence of intelligence and neurological disorders, which holds tremendous promise for advancing our understanding of both biological and artificial intelligence, and ultimately propelling the development of artificial general intelligence and facilitating precision mental healthcare.

## 1 Introduction

Demystifying the principles that account for human intelligent behaviors, such as recognizing faces and making decisions, has been attracting a tremendous amount of interdisciplinary effort and is also the driving force behind the boom in artificial intelligence. The closer we can approach the intrinsicality of intelligence, the higher the possibility that we could master the emergence of intelligence.

As the biological recesses of intelligent behaviors, the multiscale characteristics of the human brain are specifically being identified to explain the remarkable neurobiological basis underlying intelligent abilities. Even at the microscopic scale, the neuroanatomical characteristics of neurons were reported to be linked to intelligence quotients [1]. With the development of neuroimaging techniques that provide in vivo observations of the brain-wide morphological structures and functional activities, various brain regions were found to correlate with general intelligence [2,3]. Further evidence suggested that human intelligence may emerge from distributed brain activity, emphasizing the integration of information processing across brain regions [4-6]. Moreover, the interaction/causality between brain regions while conducting specific tasks is being explored to explain how the brain is involved in intelligent behaviors[7-9]. These neuroscientific findings systematically provided multi-perspective biological priors of intelligent behaviors by studying the brain along with its network organization, i.e., simultaneously perceiving the effect of brain regions and their interactions. It is also consistent with brain organization per se, which can be empirically presented as structured networks[10]. Therefore, if we want to understand the intelligent behaviors conceived in the human brain by mathematically modeling brain activity, having a systematic repository of the multiscale brain network architecture would be very useful for pushing the biological boundary of an established model.

Specifically, our point here is supported by the great success of artificial neural networks in achieving human-like intelligent behaviors, although they lack biological realism. Especially, with the unprecedented advances in computing power of graphics processing units (GPUs), highly parallelizable GPU-enabled algorithms accelerate the proliferation of deep learning techniques [11] that have been bringing revolution or renaissance in almost every field of research by way of artificial intelligence. We have recently witnessed the brilliant success of artificial neural networks in areas such as natural language processing tasks by using the OpenAI's ChatGPT [12] and computer vision tasks by using the MetaAI's segment anything model (SAM) [13]. We should also remember that the core idea underpinning these booms, i.e., the artificial neural network, was first proposed almost 80 years ago by McCullough and Pitts [14]. As they described, artificial neural networks have two basic elements including nodes and connections between nodes, which facilitate computational information flow within the network, and share a conceptually similar computational paradigm with the biological brain. Therefore, except for the rapid development of computing technologies, the current success of artificial intelligence emphasizes the importance of network architecture if our goal is to empower a model with intelligence.

Despite rising optimism that artificial intelligence will transcend biological intelligence in the near future [15], we must acknowledge that plenty of distinctions exist between the two kinds of intelligence. In specific scenarios such as playing video games, artificial intelligence can indeed obtain significantly higher performance than biological intelligence by reinforced learning from human behavioral data [16]. However, mounting evidence has demonstrated the vulnerability of artificial intelligence when facing adversarial situations that biological intelligence can easily avoid [17]. More obviously, biological intelligence can perform much more general intelligent actions, such as flexibility in shifting between tasks from different domains, than current artificial intelligence, which is usually trained with specific parameters. One of the essential reasons for these differences is that we still know too little about the computational mechanism of biological intelligence to design a more biologically plausible artificial intelligence. Whether we should heavily rely on the mechanisms of biological intelligence to achieve artificial intelligence is open to discussion. If the target is only about accurate performance in specific tasks, such as recognizing the human face, the answer may be no. But, if our goal is to have artificial general intelligence or to make the artificial intelligence behave more like human intelligence, such as flexibly switching between different tasks, we strongly suggest that we need to pay much attention to how the brain proceeds information. Therefore, to establish the connections between biological intelligence and artificial intelligence, we need a shared platform. Fortunately, as we described above, the nodes and connections-based structure of neural networks provides a common computational paradigm shared by biological intelligence and artificial intelligence. This means that we can transfer accumulated and upcoming knowledge about biological intelligence to artificial intelligence in the form of networks. But, how should we pursue this?

To accomplish this, we propose that the digital twin brain (DTB) can bridge the gap between biological intelligence and artificial intelligence in the form of brain networks. The digital twin concept was first proposed for the Apollo mission at the National Aeronautics and Space Administration (NASA) in the United States. Since then, this technology has been widely applied in manufacturing and healthcare situations by making a mockup of a physical object using real-time data to create various what-if scenarios. Ideally, if this technology can be used to represent the brain digitally, many experiments about biological intelligent behaviors and mental health care may be able to be conducted virtually to aid in increasing our understanding of the underlying mechanism. However, how the digital twin technology should be applied to simulate the brain, especially the brain functioning process, is largely unknown. Recent evidence from multiple independent studies has demonstrated the feasibility of simulating the brain at the whole-brain level. Especially, EBRAINS, which is supported by the EU-co-funded Human Brain Project, has provided The Virtual Brain (TVB) platform to accelerate full brain network simulations [18,19]. However, without enough biological plausibility, they are more useful in the field of simulation rather than as a digital twin of the brain. Therefore, we here emphasize three essential elements for establishing the DTB from the architecture to the emergence of functions (Figure 1).

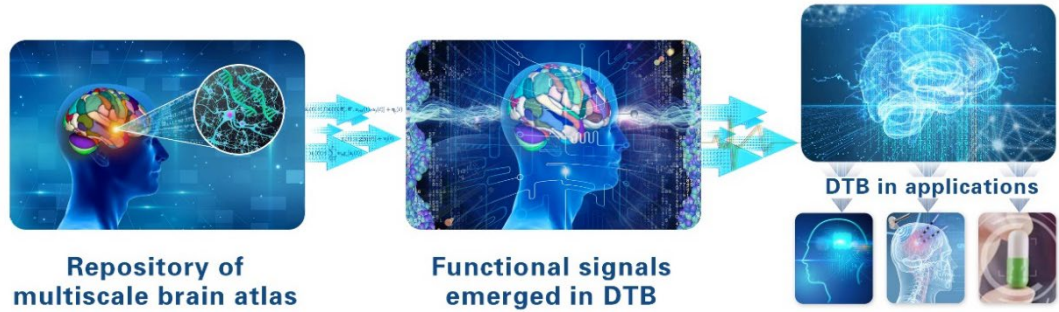

Figure 1. Three elements in the DTB and the logic relationship among them.

1) As a digital twin of the biological brain, the biological realism of DTB, reflected in its model architecture, must have close correspondence with the biological brain, such as a similar network architecture and reflecting the regional heterogeneity across the brain. Without biological realism, we can only model the brain fragmentarily, rather than digitally twinning the brain. Therefore, to seamlessly integrate the biological findings for scaffolding the DTB, a multimodal and multiscale brain atlas covering the brain network organization will be desirable.
2) When the DTB has been established by following the architecture of biological brain, it should be able to generate functional signals that are as close as possible to the biological brain by equipping the model with newly developed algorithms. That is, we should manage to train the DTB by using the collected functional data from biological brains. The result would be functionally similar to the built-in mechanisms of the biological brain in a way that emerges functional activity from the comparable fixed model structure.
3) The DTB should demonstrate its value in a broad range of scenarios, such as understanding the computing principles of brain functions and facilitating precision medicine by supporting a clinical decision with more computing evidence. By testing the DTB in different kinds of applications, we can evaluate the pros and cons of the current twin model, which would drive the evolution of the DTB with respect to the practical requirements. Finally, we might have a chance to understand how intelligence emerges from the network, how different types of brain diseases attack the network, and how the brain can leverage external stimuli to alleviate these attacks.

In addition, it should be noted that the DTB that we are proposing is far from mindclones in science fiction. Establishing the DTB is a learn-by-doing approach. That is, by modeling brain activity with biological plausibility, we hope to increase our understanding of the mechanisms behind the biological computing details, which may be incorporated to advance the current form of artificial intelligence. The rest of this paper is organized as follows. In sections 2 to 4, we will introduce the three key ingredients of DTB, i.e., the structural basis of the DTB, the three levels of models from micro single neurons to macro whole-brains to generate brain functional signals and the applications of DTB in simulating and regulating brain dynamics, separately. Section 5 will discuss the challenges and potentials of DTB for opening up new possibilities for future research.

## 2    Mapping the Human Brain: Brainnetome Atlas

Constructing brain atlases at different scales, modalities, and across different species is highly beneficial for the development of the DTB and for the computational modeling of neural systems (Figure 2). Brain atlases are comprehensive frameworks that encompass various aspects of the brain, including the demarcation of boundaries between different brain regions, characterization of their respective functions, identification of distinct neuronal cell types, and exploration of brain connectivity at different scales, ranging from macro to meso to micro levels [20-25]. This can help us understand how brain regions are interconnected and how they interact at various levels of granularity. Such insights are crucial for modeling brain dynamics and simulating complex neural processes. By training networks with biologically realistic connectivity, it can achieve superior performance compared to learned random networks [26]. Integrating data from different imaging modalities can provide a comprehensive view of brain structure, connectivity, and activity. This multimodal approach can enable us to capture complementary information about the brain, enhancing our ability to model and simulate neural activity with higher accuracy. Constructing brain atlases across different species, including humans and non-human animals, can further enable us to study evolutionary relationships and identify conserved brain regions and functional circuits [27]. This comparative approach aids in understanding the fundamental principles of brain organization and can provide insights into the neural mechanisms underlying cognition, behavior, and disease, providing significant priors for building the DTB.

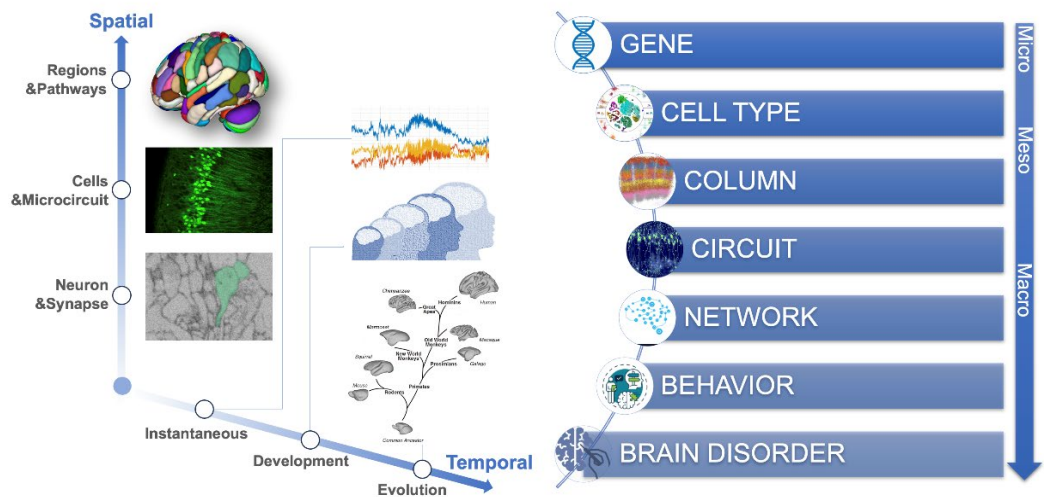

Figure 2．Multiscale and multifaceted Brainnetome atlas. Development of a multiscale and multifaceted Brainnetome atlas refers to an extensive and detailed mapping of the brain's structure and functional organization across multiple scales and dimensions. It can provide a comprehensive understanding of the brain by incorporating a wide range of information, including anatomical, physiological, molecular, and connectivity data obtained through various imaging techniques and experimental methodologies. The term "multiscale" indicates that the atlas will offer insights into the brain at different levels of resolution, ranging from macroscopic regions and networks to individual cells and molecular components. This multi-level approach will capture the hierarchical organization of the brain, allowing for a more comprehensive analysis of its structure and function. Moreover, this Brainnetome atlas will provide a holistic understanding of the brain's dynamics by considering temporal aspects, such as activity patterns, development, and evolution. By integrating temporal information, the atlas can enable researchers to explore the brain's changes across the lifespan, understand the underlying mechanisms of brain function, and study evolutionary relationships. The term "multifaceted" emphasizes that the atlas will encompass diverse aspects of brain organization and function. It will integrate multiple modalities of data, such as anatomical imaging, functional imaging, molecular profiling, and connectivity data. This multidimensional approach can enable researchers to explore various facets of the brain, including its anatomy, connectivity patterns, activity dynamics, and molecular composition.

Cross-modal, multi-scale brain imaging serves as the fundamental technology underlying the development of brain atlases. Different imaging modalities capture diverse physical and chemical characteristics of brain structure and function. It is essential to integrate these modalities to obtain a comprehensive understanding of the brain as a unified entity operating at multiple scales. Overcoming the challenges associated with integrating cross-modal information, bridging different scales, and accurately representing the spatiotemporal dynamics and intrinsic connections of brain structure and function is crucial in constructing brain atlases. Moreover, an essential endeavor is establishing the large-scale DTB. The purpose of the DTB in the form of nodes and connections-based architectures, which will be built on the foundation of a brain atlas, is to replicate the complex structural and functional characteristics of the human brain in a computational model. The integration of cross-modal, multi-scale brain imaging techniques, along with the establishment of our large-scale DTB, represents a significant step towards unraveling the mysteries of the brain.

To contribute to this, the team at the Institute of Automation of the Chinese Academy of Sciences (CAS) has proposed the further development of a human brain atlas called the Brainnetome atlas [21]. This atlas already incorporates multimodal brain connectivity information and encompasses 246 sub-regions of finely-grained brain regions, along with the structural and functional connectivity patterns between these sub-regions. By providing precise and objective localization of sub-region boundaries and elucidating their functional implications, the Brainnetome atlas has become a macro-scale atlas of whole brain connectivity in vivo. By utilizing the Brainnetome atlas, researchers can gain deeper insights into the complex workings of the brain and develop innovative approaches to address various challenges in brain science and related fields. More importantly, the Brainnetome atlas can provide a valuable resource for inspiring the development of the DTB systems, and novel design principles for brain-inspired artificial networks.

## 3  Brain atlas constrained DTB models

Computational models will be critical components of the DTB, enabling the replication of intricate characteristics of physiological signals, including raw neural activity, firing rate, fMRI, EEG, and MEG, while constrained by brain structure. To model the brain comprehensively and accurately, it is essential to consider multiple scales, encompassing the microscopic level of individual neurons, the mesoscopic level of neuron populations, and the macroscopic level of the whole brain.

At the microscopic level, neuronal models capture detailed behavior and properties of individual neurons, such as their electrical activity, ion channel dynamics, and synaptic connectivity [28]. These models, including integrate-and-fire model [29] and Hodgkin-Huxley model [30], provide insights into the fundamental building blocks of brain function and allow us to study the intricate mechanisms underlying neuronal computations. The integrate-and-fire model and its variants, have been widely used to explore brain dynamics at rest [31] and can be seen as a foundational model for artificial neural networks in that it provides a basic framework for understanding how neurons integrate and transmit information [32].

Moving to the mesoscopic level, neuron population models consider the collective behavior of groups of neurons that share common characteristics or are functionally connected [33]. These models capture the emergent properties and dynamics arising from interactions within and between populations of excitatory and inhibitory neuron. These models can be broadly categorized into two groups: biophysical models and phenomenological models. Biophysical models describe the detailed biophysical properties and mechanisms of neuronal activity, whereas phenomenological models, such as Kuramoto model that characterizes the synchronized oscillatory behavior of neural populations [34], and Hopf model that generates periodic oscillations through a supercritical bifurcation [35], focus on capturing the overall patterns and dynamics of neural activity without explicitly modeling the underlying biophysical processes [36]. Examples of biophysical models include Wilson-Cowan model that characterizes the average firing rate of neuronal populations [37], a reduced spiking network developed by Wong and Wang [38], and a dynamic mean field (DMF) model with local feedback inhibition regulating the firing rate to approximately 3 Hz for each local excitatory population by Deco et al [39,40]. By combined with multimodal data, these types of models help us understand how large-scale patterns of activity and information processing emerge from the collective behavior of neurons.

Finally, at the macroscopic level, whole-brain models integrate information from different brain regions and networks to capture the global dynamics and functional connectivity of the whole brain that are constrained by brain structure. These models enable us to investigate how different brain regions interact and influence each other, leading to complex cognitive processes and behaviors. Moreover, whole-brain models are complex systems composed of multiple factors that interact and co-evolve over time [41]. Each of these is a comprehensive framework that can be used to simulate the dynamic patterns of the brain and to investigate the underlying mechanisms that drive the observed phenomena. For whole brain modeling, model fitting is a crucial process that tunes parameters to improve accuracy, gain a deeper understanding of the underlying system, and make predictions. There are two commonly used fitting methods: parameter space exploration and model inversion. The parameter space exploration method explores all possible parameter combinations and selects the best-fit one as comprising the model parameters [42]. Model inversion is a backward method that uses machine learning to infer the posterior distribution of model parameters from observed data [43,44]. Although either method can lead to accurate results under different conditions, the types and ranges of the parameters should be chosen according to the specific context as it involves a trade-off between model complexity and accuracy. Various factors, including data preprocessing [45], the number of brain parcellations and its long- and/or short- range connections [46-48], and model paradigms [49], will affect the results to some extent; thus, there is no consensus about choosing the parameter configurations. One of the key advantages of whole-brain modeling is the ability to analyze and interpret the model's parameters in relation to real data, which enables us to gain insights into specific physiological phenomena [50]. In addition to providing a tool for studying brain functions, whole-brain models also offer a platform for testing hypotheses and predicting the response of the system under different conditions. By adjusting the model parameters, we can explore the effects of specific interventions or perturbations. This can be particularly valuable in the context of developing new therapies or treatments for brain disorders [51,52].

By considering these three levels of brain modeling, we gain a more comprehensive understanding of the brain's complexity and the networked interactions that shape its function. This multi-level approach allows us to bridge the gap between the microscopic details of neuronal activity and the macroscopic patterns of brain dynamics, ultimately advancing our knowledge of brain functions and their roles in various neurological and psychiatric disorders.

## 4  Modeling brain functions, dysfunctions, and interventions in the DTB

With the biological counterpart and bottom-layer models, there are many things that we can do within the framework of the DTB, including simulating how the brain works in the resting and task states, modeling brain dysfunctions when there are brain disorders, and restoring the brain dynamics from undesirable to target states. In the following, we review some applicational studies for the DTB from the above-mentioned three aspects (Figure 3).

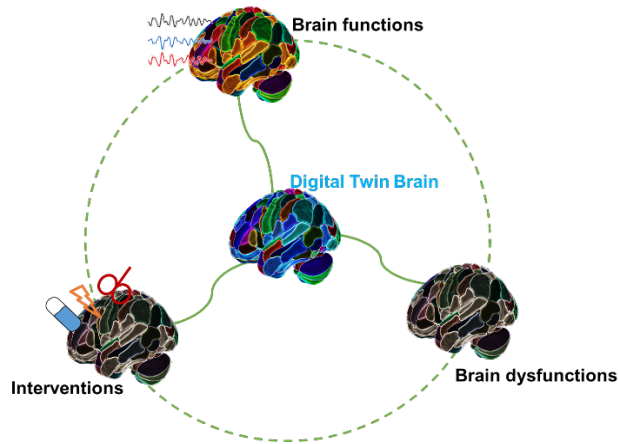

Figure 3．Schematic diagram of modeling brain functions, dysfunctions, and interventions in the Digital Twin Brain (DTB). Using DTB, we can model normal brain activities in resting states, and simulate different types of cognitive tasks. Modeling brain dysfunctions is helpful for revealing the underlying neuropathology of brain diseases by comparing it to a healthy functioning brain. Different interventions, including DBS, TMS, and pharmacological interventions, can be modeled and utilized to modulate the brain dynamics from an undesirable (e.g., unhealthy) to a desirable (e.g., healthy) state.

## 4.1 Modeling brain functions in DTB

In the biological brain, the different kinds of brain functions with specific spatiotemporal patterns are generated by a relatively stable neuroanatomical scaffold. Yet, the underlying mechanism remains unclear. The difficulty lies in the lack of a quantitative way to measure the functional computing mechanisms underlying different brain functions. By approaching the brain-like functional signals, the DTB could provide a quantitative way to understand the mechanisms through which different brain functions emerge from a consistent structural organization.

By roughly allocating different types of neurons into large brain regions, previous researchers started to simulate the human brain activity at the whole-brain level almost fifteen years ago. For example, Izhikevich et al. proposed a simplified spiking neuron model of the thalamo-cortical system that simulates one million multicompartmental spiking neurons of 22 basic types and almost half a billion synapse connections [53,54]. By offering a set of parameters, this model was able to reproduce known types of responses recorded in vitro in rats as well as to help to analyze the behavioral mechanisms of normal brain activity in humans. Similarly, based on the different functional roles of brain regions, Eliasmith et al. designed a large neural network (called the Semantic Pointer Architecture Unified Network, 'Spaun') comprising 2.5 million spiking neurons to conduct eight different tasks such as image recognition and serial working memory [55], by offering the task identifier. The same research group continued to developing Spaun 2.0 model, which comprised approximately 6.6 million neurons and was capable of performing twelve different types of cognitive tasks [56]. Although the functional modules of the model primarily had conceptual correspondence to specific brain regions, the creation of Spaun and its updated version demonstrated the feasibility of establishing models to conduct various tasks by following the organizing principles of the human brain. If the model could be improved biologically, such as by making the functional signals generated from the model become more similar to the human brain, we could gain further understanding about how the brain functions by looking into the parameter space of the model.

By modeling the functional dynamics of brain regions, several whole-brain models have been established to simulate resting-state functional connectivity at the macroscale [18,57,58]. Recently, Lu et al. set up an extremely large-scale whole-brain simulation with the leaky integrate and fire (LIF) model as the basic computing unit [59]. Specifically, this model comprised up to 86 billion neurons, close to the estimated number of total neurons in a human brain and had 10 trillion synapses for spike communication, which could simulate the brain functional activity both in the resting state and in vision and auditory tasks. However, it still has not closely touched on the ways that biological information about the brain regions could contribute to the models and how the brain functions should be generated.

Obviously, to simulate the brain function at the magnitude of a billion spiking neurons, the demands on computing resources are extremely high, e.g., 10,000 GPUs to implement simulations in the work mentioned above [59]. If more biological features such as different axonal densities and mantle layer information are added, the computational loading will likely increase exponentially. Therefore, we must seek the trade-off between biological plausibility and computational possibility. It is quite possible that accurate biological information may help to alleviate the computational loadings by merging similar neurons, for example, into cortical columns or refined brain regions. Thus, by using the Brainnetome atlas with accurate biological information, as we described above, the DTB may be started at the macroscale level. With the progress of the Brainnetome atlas into the microscale, the DTB could be updated in a more biologically meaningful direction. At the same time, we need to develop new models that can represent the biological features efficiently. Additionally, new neuromorphic computing hardware, such as BrainScales [60], Pohoiki Springs [61], and Darwin Mouse [62], may contribute to accelerating the large-scale twinning process with lower computational costs.

Along with these studies on functional simulation at the whole-brain level, more studies were made from the perspective of computational neuroscience to investigate the dynamic mechanism of brain functions based on the local neural circuit. For example, the spiking circuit model has been applied to investigate the neuronal mechanism of decision making [38,63,64] and the emergence of stable neuronal timescales to support working memory [65]. These recognized local circuits and dynamic principles provide a repository for simulating the local activity of specific brain regions when establishing the DTB for the corresponding tasks. However, more effort is still needed to establish a series of methods that can interrelate these local circuits and fit them into the model efficiently.

## 4.2 Modeling brain dysfunctions in DTB

Because they are affected by various factors including physiology, psychology, and social environment, the underlying pathogenic mechanisms of many psychiatric or neurological disorders remain elusive, thus requiring new techniques in addition to experimental studies and theory. In recent years, computational models have provided a new perspective for the mechanism analysis of brain functions and its disorders [66], such as schizophrenia [67,68], brain tumors [69,70], and epilepsy [71]. The advantages of this way of doing research lie in the fact that these methods not only can simulate bio-plausibly dynamic mechanisms of brain diseases at the neuronic scale, at the level of neural populations, as well as at the brain region level but can also perform virtual surgical treatments that are impossible to perform in vivo due to experimental or ethical limitations [72]. These methods provide powerful tools for the study of brain diseases via the relationship between brain structure, function, and dynamics. Here, we provide an overview of some influential studies from the two perspectives of functional abnormalities and structural abnormalities.

Studies have shown that many mental or neurological diseases are related to connectivity imbalances. Altered functional networks in these diseases are thought to be related to changes in excitatory-inhibitory (E/I) imbalances in local microcircuits [73], which can be simulated by biophysical models. For example, schizophrenia (SCZ) is often conceptualized as a disorder of altered brain connectivity [74] and is hypothesized to be associated with elevated E/I ratios in cortical microcircuits [75]. Yang et al. utilized a dynamic mean-field model to study the dynamics of brain networks as a function of local and global parameters and found that the empirically observed increases in voxel-wise variability and global signals in SCZ might arise from local recurrent self-coupling within nodes and long-range, global coupling between nodes [68]. The same group extended this line of research to simulate the impact of elevated E/I ratios on model-derived functional connectivity by alternating key biological parameters [67]. This study revealed that the model predicted FC hyperconnectivity for elevated E/I ratios and showed that models that account for the heterogeneity of association and non-association cortical regions could better explain the spatial pattern of functional connectivity changes observed in neuroimaging data in SCZ. Benefiting from computational models, such disruption of macroscopic brain connectivity can be explained. In addition, such research on brain dysfunction caused by E/I imbalance in local microcircuits has also been widely applied to brain diseases, including Alzheimer's disease [76] and stroke [77].

Some other brain diseases arise from disruptions in brain functional connectivity caused by alterations in structural networks, and their effects on brain activity can generally be investigated by simulating lesioning (removing edges) or resecting (removing nodes) specific brain structures [78]. Based on personalized structural connectivity, Aerts et al. analyzed the impact of changing structural connectivity on a postoperative spatial function by using "virtual neurosurgery" for a tumor resection, and found that this method improved the prediction of postoperative brain dynamics in patients with brain tumors [69,70]. Likewise, to assess the impact of lesions on temporal dynamics, Wei et al. simulated the lesion of a specific node by regulating the connection strength of specific nodes [79]. This work revealed that lesion effects exhibited regional dependence and could be predicted by the anatomical hierarchy axis and specific network measures of structural brain networks.

Specially, epilepsy is a disease known to be associated with both structural and dynamic changes in the brain, often spreading from an onset zone to other distal areas along white matter tracts [71,80]. From the perspective of macro-scale, many leading studies of epilepsy so far have come from the group of Jirsa et al. [43,71,80,81]. They proposed a virtual epileptic patient (VEP) based on individual brain networks to simulate the individual seizure propagation patterns [71]. In this model, whether each node is an epileptogenic zone can be determined according to the excitability threshold. The model was applied to clinically useful estimations with promising results by comparing the predicted foci with clinical decisions [82-84]. In addition to studies such as these from the macro-scale perspective, there are many other computational models of epilepsy, including ones on a single-neuron scale [85].

Collectively, computational models have the potential to provide insights into the mechanisms of brain diseases and aid in the design of interventions. It is worth noting that, in addition to the above-mentioned mean field models for brain diseases, some single-neuron models also have a wide range of applications in brain diseases, such as in modeling striatal microcircuits based on the single-compartment models with Hodgkin–Huxley-type dynamics to simulate potential source of the enhanced beta rhythms in Parkinson's disease [86]. Critically, applying anatomical connectivity as a structural scaffold requires authentic and reliable brain atlases, and incorporating prior knowledge of the brain, such as spatial heterogeneity/hierarchy characterized by intracortical myelin content [87], functional gradient [88], and gene expression profiles [89], into the computational model would provide a reasonable improvement for the mechanism research of brain diseases. In addition, disease-specific models should be formed according to the specific etiology of the disease, since different brain disorders have different pathologies, as reviewed here and elsewhere [72]. These models would also benefit from iterative updating along with increased information of brain dysfunctions.

### 4.3 Modeling interventions of brain diseases in DTB

The goal of modulating brain dynamics is to help predict the outcomes of an external intervention, either physical or chemical, as well as to design new therapeutic strategies for brain diseases. In the current era of the rapid development of intelligent computing, modeling interventions of brain diseases in DTB has the potential to provide cost-effective evaluations for clinical diagnosis and treatment before beginning clinical trials. Currently, intervention techniques, including invasive deep brain stimulation (DBS), non-invasive brain stimulation (NIBS) and pharmacological intervention, are widely modeled to study the resulting brain dynamics at multiple topological scales from a phenomenological or predictive perspective [52,90].

#### 4.3.1 Modeling invasive neuromodulation: DBS

Deep brain stimulation (DBS) is a typical invasive technique that induces modulation by applying high-frequency electrical stimulation to a specific brain area and is an effective physiotherapeutic method for neurological and psychiatric disorders, such as Parkinson's disease (PD) [91] and treatment-resistant depression [92,93].

Using PD as a case study of the application of DBS, attempts have been made to model neuronal dynamics in PD from the microscopic scale of single neurons to the macroscopic scale of whole-brain networks [94]. Biophysical modeling of PD has mainly focused on the cortex-thalamus-basal ganglia circuit through direct, indirect, and hyper direct pathways of projections from the cortex to subcortical nuclei, and the models have included the conductivity-based Hodgkin-Huxley model, simplified spike-based Izhikevich model, or the reduced mean-field model [95]. Key clinical biomarkers of PD, such as pathological electrophysiological oscillatory (e.g., beta band) activity, were reproduced to reveal the potential mechanisms of action of DBS [96] and further to identify the optimal stimulus parameters to restore normal neuronal dynamics at the right stimulation frequency [97] and time [98].

In most of thalamo-cortical microcircuit models, the entire cortex was modeled as a single spiking network node, and simple synaptic connections that generate neural activity were modeled. However, neuromodulation-induced changes in cognition do not arise directly from the modulation of individual neurons, but by neural populations and circuits at the mesoscopic level [99]. In addition, the highly networked structure of brain implies that localized perturbations not only yield localized effects but also induce indirect effects that propagate along neural pathways [100], making it more appropriate to simulate neuromodulation from the meso- and macroscopic scales. Therefore, a whole-brain macroscopic perspective for understanding the networked modulating effects of DBS is needed. Recently, a large-scale network model was proposed to bridge the microscale of single spiking neurons and whole-brain signals in a multiscale model for the purpose of virtually performing stimulations and forecasting the outcome of DBS for PD patients [101]. From the macroscopic scale, whole-brain network model was used to uncover the biophysical mechanism and to explore stimulation targets for major depression [102] or combined with network control theory to design the optimal control strategy [103,104]. However, in most of the computational DBS studies, the electric field potential (voltage distribution) induced by the stimulation was not considered, which is useful but less realistic. Usually, the stimulus-induced potential field is forwardly calculated by a quasi-static approximation and the finite element method [105]. Recent progress has been made on a large-scale computational model that characterizes the spatiotemporal response following DBS for treatment-resistant depression at the whole-brain level when considering the electric field potential surrounding the active contact position, showing its capacity and potential for identifying stimulation sites and parameters [106]. Yet, the computational integrations of DBS-induced electric field and large-scale models remain to be further developed.

**4.3.2 Modeling non-invasive neuromodulation: tES and TMS**

Non-invasive brain (electrical) stimulations (NIBS) modulate brain functions by applying an electrical current to the scalp. These electrical stimulations include the widely used transcranial magnetic stimulation (TMS) and transcranial electrical stimulation (tES), which includes transcranial alternating current stimulation (tACS) and transcranial direct current stimulation (tDCS) [90]. The main difference between TMS and tES is that TMS generates a magnetic field that induces an electric field that activates neurons, while tES modulates neural activity without inducing action potentials using low current strengths [105,107].

The computational modeling of NIBS is more complicated than DBS as NIBS is an indirect neuromodulation [108]. Usually, the stimulus-induced potential field of NIBS is calculated with high-resolution head models and then used as input to cortical neurons that simulate neural responses. This types of integrative modeling was studied on the microscopic level of multi-compartmental neurons, for computational estimations of TMS and tES induced neuronal responses [105]. However, on the mesoscopic and macroscopic levels, the interaction of external fields with neuronal models are barely considered in the modeling. In most biophysical modeling of NIBS, the TMS- and tES-induced neural responses were phenomenologically modeled to reproduce experimental data and then provide mechanistic explanations, such as how tACS entrains alpha oscillation in the thalamo-cortical system [109] and how tDCS [110] and tACS [111] influence the spatiotemporal dynamics of cortical network dynamics, using spiking neuron models without considering an externally stimulus-induced potential field. Similarly, there are also many research studies that focused on the modeling of the motor cortex for the reproduction of indirect responses [112] and motor-evoked potentials [113,114] following TMS. The underlying mechanism of TMS-induced neural plasticity was also computationally explained by neural population models based on neural field theory [115,116] while neglecting the impact of an externally-applied field.

The NIBS typically stimulates a few square centimeters of cortex; thus, it is more appropriate to simulate the collective dynamics of neurons at the macroscopic scale [116]. Meanwhile, the current or voltage distribution induced by an externally-applied field, identified through a forward calculation, is critical for the biophysically plausible modeling of NIBS at the macroscopic scale. One study applied the tDCS-induced current density distribution in a whole-brain model to investigate how tDCS effectively changed the spatiotemporal dynamics in resting state functional connectivity [117], making a step forward toward biophysically realistic modeling. Despite progress, like the DBS studies, how to incorporate neuronal ensemble dynamics and an externally-applied field is still a critical issue to be addressed for the predictive large-scale modeling of NIBS.

### 4.3.3 Modeling pharmacological intervention

As mentioned previously, many neuropsychiatric diseases are hypothesized as having an E/I imbalance. Pharmaceuticals are designed to restore this E/I imbalance in the brain, by exciting or inhibiting the binding of related neurotransmitters' receptors. For example, 5-hydroxytryptamine psychedelic can bind with serotonin receptors to generate modulating currents; ketamine and other NMDA receptor antagonists can directly bind with NMDA receptors to affect excitatory neural currents. While modern biotechnology usually focuses on understanding the metabolic pathways related to disease states and manipulating these pathways using molecular biology or biochemistry, the DTB is designed to simulate changes in brain states caused by different neurotransmitters or neurotransmitter receptor alterations, shedding new light on which and how substances regulate brain functions from a computational perspective; the focus will be especially on neuropsychiatric disorders, such as depression [118], Alzheimer's disease [119], and addiction [120,121].

At present, many studies have explored the quantitative relationship between the neurotransmitter concentration and the modulating current in vivo [122]. There are also models that describe the effects of neurotransmitters, such as dopamine and $\tau$ protein at different concentrations, on the E/I balance [41,123,124]. But studies that directly investigate the impact of pharmaceuticals on the E/I balance at the macro scale are still in their infancy. Here, we list some related studies in Table 2 and introduce a dynamic coupling model for psilocybin to demonstrate the potentials of pharmacological intervention using the DTB. Psilocybin, a type of 5-hydroxytryptamine type psychedelic, is an agonist of the brain's serotonin 5-hydroxytryptamine 2A (5-HT$_{2A}$) receptor, and 5-HT psychedelics-assisted psychotherapy has demonstrated significant positive relief of anxiety and depressive symptoms in patients with psychological and social distress [125-128]. To explain the functional effects of serotoninergic 5-HT$_{2A}$ receptor stimulation with psilocybin in healthy humans, Deco et al. simulated the release-and-reuptake dynamics of the 5-HT$_{2A}$ neurotransmitter system by inducing the 5-HT$_{2A}$ receptor density and coupled it with the neuronal activity at the whole-brain scale [118]. This mutual coupling model gives new insights into how psilocybin acts on the serotonin system and further modulates the brain activity, and, thus, shows considerable promise as a therapeutic intervention for depression. In short, the impacts of chemical substances on neural activity can be predicted prior to clinical trials, which will help in developing drugs, designing pharmacological interventions, and providing guidance for brain diseases to rebalance human brain activity in silico utilizing the explanatory and predictive power of the DTB.

Table 1. List of computational studies in pharmacological interventions.

| References | Substance | Model | Model fit to | Parcellation |
|---|---|---|---|---|
| Deco et al. (2020) [118] | Psilocybin | Dynamic mean field model | Kullback-Leibler distance | AAL (90) |
| Deco et al. (2018) [129] | Lysergic acid diethylamide (LSD) | Dynamic mean field model | BOLD FC | AAL (90) |
| Jobst et al. (2021) [130] | LSD | Hopf model | BOLD FC | AAL (90) |
| Ruffini et al. (2022) [131] | LSD | Ising model | Ising temperature et. | AAL (90) |
| Stefanovski et al. (2019) [119] | NMDA receptor antagonist | Jansen-Rit model | Neural activity EEG | Glasser (360) |
| Coronel-Oliveros (2021) [120] | Nicotine | Jansen-Rit model | Functional Connectivity Dynamics | AAL (90) |
| Coronel-Oliveros (2023) [121] | Nicotine | Jansen-Rit model | Functional Connectivity Dynamics | Schaefe (100) |
| Deco et al. (2011) [132] | Acetylcholine | Leaky Integrate-and-Fire model | Neural activity | / |
| Sajedin et al. (2019) [133] | Acetylcholine | Leaky Integrate-and-Fire model | Neural activity | / |

## 5 Perspectives

The DTB opens a new avenue for elucidating the basic principles of the brain and for revealing the complex neural mechanisms behind brain functions and dysfunctions, establishing a new paradigm of integration between the understanding, simulating, and controlling of the brain, and more importantly, bridging the gap between biological intelligence and artificial intelligence. As reviewed above, significant progress has been made so far. However, it is in its infancy, and many open questions remain. Here we present some representative ones that are essential for the study of DTB.

**Achieving a digital twin of the biological brain based on a cross-modal, multiscale brain atlas**

The core idea lies in how to draw inspiration from and extract the framework and dynamic information provided by the brain atlas and develop relevant modeling methods to make a digital twin of the brain. This can be achieved from the following two perspectives. First, the development of the brain atlas provides a real data foundation for establishing a digital twin of the biological brain, including the reference of structures and functions at different spatial and temporal scales. This ensures that the digital twin of the brain will approximate the real neural system, not only in terms of structural resemblance but also in terms of functional similarity. Second, in the improvement and validation of cognitive computing models, the brain atlas will provide multi-omics, multiscale structural descriptions, and dynamic activity information. This includes genetic information related to the genome, which is currently lacking in artificial intelligence models, the coupling and generative relationships between brain structure and function, knowledge of conservation and variation in the evolution of the nervous system, as well as the low-dimensional and nonlinear basic laws of the nervous system in terms of structure and function.

The knowledge provided by the brain atlas and the latest biological discoveries will continuously improve the digital twinning of the brain, making it more and more realistic, as the recent progress that has been made in simulations informed by brain structure [87], function [88], and genetic profiles [89] has already shown. These studies have shown that the multiscale and heterogeneous organization of the biological brain is crucial for building a bio-plausible DTB and understanding the structure-function-dynamics relationships of real brains, thus linking micro-meso-macroscopic phenomena with behavior. Based on multimodal brain imaging and multi-omics data, relating macroscopic phenomena and behaviors to their mesoscopic circuits and microscopic neurons or synapses is critically vital for future DTB research. Ultimately, this advancement will drive the realization of general artificial intelligence and the understanding of the source of intelligence.

**How to build a more realistic and reliable model?**
A more realistic model of the brain is crucial for advancing our knowledge and applications in neuroscience. Such a model enables us to delve deeper into the complexities of brain function, make more accurate predictions, and improve clinical interventions. Although computational models of the brain have made remarkable progress in various applications, there are still several challenges and constraints that still need to be addressed.

- When modeling a single neuron, it is crucial to consider how to effectively capture the spatiotemporal characteristics beyond just describing the electrical activity of dendrites [32]. Neurons exhibit intricate branch structures and complex spatial topologies. Existing models often either focus solely on depicting synaptic membrane potentials, capturing only temporal dynamics, or simplifying neurons as one-dimensional entities, as in the case of cable equations [134]. To enhance our understanding of information processing and propagation in the brain, it is crucial to advance modeling approaches that embrace multi-dimensional space. By incorporating multi-dimensional space modeling, we can more accurately capture the intricate shapes and connectivity patterns of neurons. This, in turn, can enable us to gain deeper insights into neural phenomena such as synchronization, oscillation, and network dynamics. A comprehensive exploration of multi-dimensional space modeling will contribute to advancements in neuroscience research and our understanding of the complexities of neural systems.
- When modeling neuron populations, a critical issue is how to effectively integrate diverse structural components to enhance neuronal activities and capture cross-scale interactions. This includes incorporating factors such as connectivity [31], geometry [135], and microstructure [136]. The current cross-scale integration models have made notable advancements in nesting micro-meso-macro layers [118]. However, their focus is often limited to individual or multiple microscopic attributes [41], failing to capture their full mutual influence and interactions. To propel these models forward, a critical next step is the comprehensive integration of various microscopic attributes, including genetic factors, receptor dynamics, myelin integrity, and more, with a particular emphasis on characterizing their interactions. By incorporating these diverse elements into the models, we can achieve a more comprehensive understanding of the complex interactions and dynamics within the neural system, leading to insights into brain function and information processing at multiple scales.

- When modeling the whole brain, how to verify the model's accuracy and reliability is a key point in the transition from theory to application [83]. While models aim to simulate real-world dynamics as closely as possible, it is important to remember that simply fitting the data well does not guarantee the usefulness or validity of the model. While significant progress has been made in whole-brain simulation, with a primary focus on mechanistic explanations and simulation regulation, the transition to practical applications necessitates validation with real data. The next critical step involves rigorous testing and validation of these simulations using empirical data and experimental findings. This will ensure the accuracy, reliability, and relevance of the models in real-world scenarios, enabling their successful integration into practical applications, such as clinical diagnosis, treatment planning, and personalized medicine. By bridging the gap between theory and application, whole-brain simulation can revolutionize neuroscience research and improve our understanding and management of brain-related disorders.

**How can DTB have multimodal human-like intelligence?**
By using Brainnetome priors that emphasize the integration of the multiscale network-based biological priors of the human brain organization, the new architecture of DTB will be able to be designed with greater biological fidelity. However, great challenges still exist in combining or hybridizing human intelligence and DTB from low-level perceptual functions (e.g., visual and auditory processing) to high-level cognitive functions (e.g., language and memory).

The most obvious difference between the human brain and the DTB may be the inputs. The human brain can easily process multimodal inputs, such as visual, acoustic, tactile, and olfactory ones. Without comparable inputs, subsequent processing by the DTB may then not be comparable with human brain functioning on the multimodal inputs. Therefore, finding a way for the DTB to receive multimodal and human-comparable inputs would be the first challenge along the way to approaching human intelligence. One possible way may use current deep learning techniques to generate human-comparable inputs. Accumulating evidence has demonstrated that convolutional neural networks (CNNs) have activity that is similar to the human functional activities of the visual cortex when processing visual inputs [137,138]. Similarly, CNNs have demonstrated correspondence with the functional activities of human brain when encountering the same acoustic stimuli [139-141]. Therefore, pretrained neural networks may be adopted to provide modal-specific input to the DTB. However, it would still require further effort to validate and harmonize these inputs.

Another key property of human intelligence is the functional flexibility that can be seamlessly achieved in the same system. This actually is a huge challenge and ambition for Artificial General Intelligence (AGI). The current path to AGI is mainly via large AI models, such as large language models consisting of more than 100 trillion parameters. By increasing the parameter space, large models may be able to have state-of-the-art flexibility in specific task domains. However, the human brain does not achieve its functional flexibility by having specific large models for specific task domains. In contrast, in a diverse task context the human brain can respond smoothly to multiple inputs while using the same biological system. Therefore, although this would be a further challenge about how the DTB could achieve functional flexibility, it might also bring us the opportunity to establish small but efficient AGI models and further evidence for "whether or to what extent intelligence comes from brain organization". A promising approach may be to consider the neuromodulatory system in the subcortical nuclei, which relays and gates the flow of sensory information to the cortex for adaption to diverse inputs. The connectivity patterns between the subcortical nuclei and the cortex can be learned, which may further be adopted to design the gating components of the DTB for conducting various tasks. Surely, a plethora of further research studies are strongly needed.

Ultimately, we would like to have not just an intelligent supercomputer from the brain-inspired artificial neural network but also to enhance the understanding of the mechanism of human intelligence through brain-organization-aware models. It would then be a virtuous cycle to decipher human intelligence by modeling it using the known brain priors and then to further refine the model using the newly learned insights.

**DTB in brain diseases: from in silico to clinic**
Studies of brain diseases can obviously benefit from the DTB by analyzing the mechanisms underlying the disease and by simulating clinical surgery, brain stimulation, drug therapy, etc. Currently, the neuropathologies of many brain diseases are not fully understood, but new findings keep popping up [142-148]. This will make the DTB an invaluable tool for exploration and validation. More promisingly, it is possible to use DTB to investigate different hypotheses about brain dysfunctions. We can also manipulate brain activities in silico in unprecedentedly rich ways that cannot be performed in real human brains, potentially leading to results of great scientific and clinical significance.

The DTB has the potential to provide a new paradigm for neuromodulation by virtually simulating both physical and chemical interventions addressing changes in brain dynamics. Taking the treatment of major depression as an example, a major challenge of traditional treatment methods, including brain stimulations (e.g., DBS) and pharmacological treatments (e.g., ketamine), is to test the effectiveness of these treatment modalities via preclinical experiments in humans [149], which are usually not allowed in the real world because of ethical or other constraints. Using DTB, potential treatments for major depression may be able to be harmlessly and effectively tested from a variety of perspectives. Such testing could include safe drug dosages of ketamine and optimal stimulation targets of TMS or DBS. Moreover, most of current medicines targeting neural systems have limited efficacy in spite of a long development cycle. The DTB has potential for use in drug discovery, for example, by simulating possible effects and side-effects of newly developed substances to screen for potential drugs. By using model support from DTB, we might be able to guide and largely shorten the process of drug discovery, empowering more efficient drug development. To achieve this, we need to properly simulate the way that these chemical substances act on neural systems and how the neurochemical systems interplay. Linking microscale molecular dynamics or pharmacodynamics and system-level brain dynamics might be a possible way to model pharmacological interventions with biochemical plausibility. This new paradigm can also be similarly applied to brain stimulations, and the stimulation configurations, e.g., optimal targeted brain area and stimulation frequency, can be explored and assessed in the DTB. Nevertheless, as mentioned before, how to integrate brain dynamics with an externally-applied field is still a critical issue for both DBS and NIBS and needs to be addressed before we can obtain biophysically realistic neurostimulations.

There are individual differences, and diagnosis and treatment are made on a case-by-case basis in clinical contexts. In particular, it is quite likely that specific brain areas or circuits need to be modulated during specific states for different patients in neurostimulation scenarios. Applying DTB has the potential to eventually facilitate precision medicine for brain diseases digitally, but this requires modeling at the individual level although current studies are mostly group-wise or use individual features (e.g., structure connectivity) that have been extracted from a group template. Thus, it will be necessary to build an individualized DTB that can enable personalized virtual therapies. But how? An individualized brain atlas might provide the basis for adapting the DTB to the individual level. Another possible approach to individualization, inspired by the idea of pretraining, is to pretrain the DTB on a large sample of healthy controls and then fine-tune it using the personalized data of a specific patient.

In general, efforts are still underway to advance the study of brain diseases in DTB from phenomenological to predictive modeling and further from in silico to clinic. The study of brain diseases using the DTB will require applying information from cross-modal and multiscale brain atlases, gaining new knowledge about brain dysfunctions, and designing bio-plausible assumptions-based models. This is a challenging process but would greatly benefit the design of personalized, efficient, and effective treatment strategies for a variety of brain diseases in the future. Meanwhile, we should always be aware that these computational frameworks must be carefully validated with neurobiological experiments and empirical data before using the DTB in clinical applications.

**Key challenges of the DTB platform**

The realization of cross-modal and multi-scale DTB simulation cannot be achieved without the support of an efficient simulation platform. Building such a platform involves multiple technical fields, including numerical simulation and modeling, neuroinformatics, neuronal and neural network modeling, brain dynamics modeling, brain atlases, as well as front-end and back-end engineering development based on different technology stacks, algorithm optimization, multi-core parallelism, and hardware acceleration using GPUs. There are a few neuroinformatic platforms, which cover the brain from the single neuron to the whole brain. These include Neuron [150], Brain 2 [151], BrainCog [152], BrainPy [153], The Virtual Brain [18,154], and Neurolib [155]. However, most of them suffer from limitations that include: 1) There is a lack of support for the modeling constrained with a brain atlas that can serve as the foundation for building digital brains. Relevant atlases are either not integrated into existed platforms or are available in limited quantities. 2) Cross-modal multiscale simulation capabilities are limited. 3) They have low simulation efficiency due to poor code optimization or inadequate support of hardware acceleration. 4) These often have poor user-friendliness, but researchers often require a friendly interactive interface for model-building and simulations. Current platforms, however, primarily rely on coding, making them difficult to use. 5) There are few visualization and analysis modules that can make the simulation processes and results easily visualized in a timely manner. This also does not enable users to conduct rapid quantitative analyses of the simulation results so that the models are limited in their ability to provide a one-stop modeling, simulation, visualization, and analysis function for scientific researchers. In brief, the current neuroinformatic platforms cannot readily support multiscale cross-modal DTB modeling and simulations that are based on a brain atlas. Therefore, it is crucial to develop an open-source, efficient, flexible, user-friendly brain atlas-constrained DTB platform that supports multiscale and multimodal modeling.

In conclusion, research studies that can be used to build the DTB require collaborative efforts of researchers from different scientific backgrounds. Previous international projects and initiatives, such as the Blue Brain Project [156], the Human Brain Project [157], the BRAIN Initiative [158], and the Neurotwin Initiative [159], have been launched and have made fruitful achievements, beginning the step toward development of the DTB. These types of collaboration will hopefully accelerate the realization of our DTB, which will bridge the gap between biological intelligence and artificial intelligence, fostering the development of artificial general intelligence and precision medicine.

**Acknowledgments**
This work was supported by the STI2030-Major Projects (Grant No. 2021ZD0200200 to T.J.), National Natural Science Foundation of China (Grant No. 82151307 to T.J.), Science Frontier Program of the Chinese Academy of Sciences (Grant No. XDBS01030200 to T.J.), and the Key Research Project of Zhejiang Lab (No. 2022KI0AC02 to T.J., No. 2022ND0AN01 to T.J.). The authors also thank Prof. Yu Zhang, Prof. Jian Cui, Zhichao Wang, and Chen Wang for their constructive suggestions that improved the manuscript.

**Competing interests**
The authors declare no competing interests.